\documentclass[12pt,draft]{article}

\usepackage{amssymb}
\usepackage{latexsym}

\newcommand{\N}{{\mathbb N}}

\newcommand{\R}{{\mathbb R}}
\newcommand{\dd}{{\rm d}}
\newcommand{\dv}{{\rm div}}
\newcommand{\tr}{{\rm tr}}
\newcommand{\supp} {{\rm supp}\:}

\font\fone=cmr10 scaled\magstep3
\font\ftwo=cmr7 scaled\magstep3
\begin{document}
\thispagestyle{empty}
\vspace{1.5in}

\centerline{\fone A Rigidity Theorem for non-Vacuum Initial Data}
\vspace{ 0.3in}

\centerline{\ftwo G\'abor Etesi}
\vspace{0.1in}
\centerline{Yukawa Institute for Theoretical Physics,}
\centerline{Kyoto University,}
\centerline{606-8502 Kyoto, Japan}
\centerline{\tt{etesi@yukawa.kyoto-u.ac.jp}}
\vspace{0.2in}

\begin{abstract}
In this note we prove a theorem on non-vacuum initial data for general
relativity. The result presents a ``rigidity phenomenon'' for the
extrinsic curvature, caused by the non-positive scalar curvature.

More precisely, we claim that in the case of an asymptotically flat
non-vacuum initial data if the spatial metric has everywhere non-positive
scalar curvature then the extrinsic curvature cannot be compactly
supported. 
\end{abstract}
\centerline{Keywords: {\it Initial value formulation, extrinsic
curvature}}
\centerline{PACS numbers: 04.20.Ex, 04.20.Ha}
\pagestyle{myheadings}
\markright{G. Etesi: Rigidity of the Gravitational Field}

\section{Introduction}
According to our experiences there are many different
gravitational configurations in our physical world. Therefore if general
relativity is a correct theory of gravitational phenomena (at least at low
energies) then it is important to know whether or not these various
patterns can be modeled in general relativity i.e. Einstein
equations provide enough solutions for describing many different
gravitational fields. Unfortunately or fortunately Einstein equations form
an extraordinary difficult system of nonlinear partial differential
equations for the four dimensional Lorentzian metric mainly because of the
rich self-interactions of the gravitational field; hence in general it is
a hard job to write down explicit solutions in this theory. Therefore
all methods which prove at least the existence of solutions are 
very important. From this viewpoint, the Cauchy problem or
initial value formulation of general relativity is maybe the most powerful
method to generate plenty of solutions.
 
As it is well-known, the initial value formulation gives rise to 
a correspondence between globally hyperbolic space-times and
gravitational initial data. Maybe we can say without an exaggeration
that the class of globally hyperbolic space-times is the most important
class of space-times from the physical point of view. Consequently the
initial data formulation provides not only many but also physically
relevant solutions. The constraint equations between initial data 
are in the focal point of the initial data formulation. The question is
whether or not these constraint equations are easier to solve than the
original Einstein equation itself making the method effective. Of course,
the answer is typically yes. 

This motivates the serious efforts made in order to understand the
structure and provide solutions of constraint equations. Far from being
complete we just mention the early works of Lichnerowicz \cite{lic1}, 
Choquet-Bruhat et al. \cite{bru}\cite{cho-ger}\cite{cho-yor} ,
Fisher-Marsden \cite{mar-fis}, Christodoulou--Kleinerman\cite{chr}. These
works mainly deal with the analytical properties of the
solutions. Witt proved the existence of solutions on a general
three-manifold \cite{wit}. More recently, in a sequence of papers   
Isenberg, Moncrief, Choquet-Bruhat and York proved the existence of
solutions under milder and milder assumptions, cf. e.g.
\cite{ise}\cite{cho-ise-mon}\cite{ise-mon}\cite{cho-ise-yor}.

The constraint equations involve the scalar curvature of the metric on the
underlying Cauchy surface which is a three dimensional smooth manifold.
Various properties of the solutions depend crucially on the scalar
curvature, especially on its sign. But we know
that in the problem of describing the sign of the scalar curvature,
especially on a compact manifold, one encounters with the topology of the
space. Parallelly to the investigations of solutions of the constraint
equations by physicist and mathematicians, mathematicians proved
remarkable results on the properties of the scalar curvature of Riemannian
manifolds. By an early general result of Kazdan and Warner \cite{kaz-war}
we know that for compact manifolds of dimension greater than two there is
no constraint on the scalar curvature {\it if there is at least one
point where it is negative}. This shows that it is easy to construct
manifolds with negative scalar curvature. If we wish to construct
manifolds with non-negative scalar curvature, however, we have to face
various obstacles coming from the topology of the manifold. We just
mention two basic examples. By results of Lichnerowicz and Hitchin, on
spin manifolds it is often impossible to construct metrics of positive
scalar curvature because of a subtle topological invariant, the so-called
$\hat{A}$-genus \cite{hit}\cite{lic2}. Moreover, in three dimensions, the
size of the fundamental group provides another obstruction for positive
scalar curvature by results of Gromow--Lawson \cite{gro-law} and
Schoen--Yau \cite{sch}. An excellent survey on this branch of differential
geometry is \cite{bes}. 

These observations make it not surprising that the topology of the Cauchy
surface has a strong influence on the properties of initial data on
it. The goal of this paper is to understand this link a bit better. Our
motivation is a paper by Witt \cite{wit} who studied the problem of
existence of initial data on general three manifolds and examined certain
characteristics of these initial data. In Section 3 we prove a theorem
which states that on open Riemannian manifolds with everywhere
non-positive scalar curvature the extrinsic curvature field of a
non-vacuum initial data cannot be compactly supported i.e. it has a
``tail'' at infinity although this tail may have sufficiently fast
fall-off to make such an initial data still asymptotically flat. The
proof of this theorem is elementary and is based on the
following idea.

By using the initial data set $(M,g,k)$ and the assumption that $\supp k$
is compact, we construct another ``universal''
initial data set $(M,g, \varphi g)$ where $\varphi :M\rightarrow\R$
is a compactly supported at least once continuously differentiable 
(or $C^1$-) function on $M$ (with a little more effort this function could
be smoothened but we do not need this). However this leads us to a
contradiction if the scalar curvature of $g$ is non-positive
everywhere. In other words, we deform the original initial data set into a
standard one whose properties are easier to understand.
\vspace{0.1in}

\noindent{\bf Acknowledgement}. The author thanks Prof. H. Kodama
(YITP, Japan) and Prof. L.B. Szabados (KFKI, Hungary) for the fruitful
discussions and to the unknown referee of JMP for calling our attention
to some weak points in the original version of the proof.
The work was supported by JSPS grant no. P99736.

\section{Background material}
First let us introduce some notations. Let $W$ be a smooth manifold. 
We will call a tensor field $T$ of type $(m,n)$ over $W$ if it is a smooth
section of the bundle
\[T^{(m,n)}W:=\underbrace{TW\otimes ...\otimes
TW}_m\otimes\underbrace{T^*W\otimes ...\otimes T^*W}_n.\]

Remember that an {\it initial data set for general relativity} is a triple
$(M, g, k)$, where $M$ is a (not  necessarily compact) connected,
oriented, smooth three-manifold,  $g=(g_{ij})$ is a smooth, complete
Riemannian metric on $M$ i.e. a non-degenerate smooth symmetric tensor
field of type $(0,2)$ on $M$ while $k=(k_{ij})$ is a smooth, symmetric
tensor field on $M$ also of $(0,2)$-type. These fields must satisfy the
following constraint equations \cite{haw-ell}\cite{wal}:
\begin{equation}
\left\{ \begin{array}{ll}
 s_g -\vert k\vert_g^2 +\tr^2k=16\pi\rho ,\\
\\
\dv (k-(\tr k)g) =8\pi J, \\
\\
\rho\geq\vert J\vert_g\geq 0.
\end{array}
\right.
\label{kenyszer}
\end{equation}
Here $s_g$ is the scalar curvature of the metric $g$ and
$\vert\:\cdot\:\vert_g$ denotes various norms given by the induced
scalar product on $T^{(m,n)}M$, e.g. $\vert k\vert_g^2 =
\left< k,k\right> =k_{ij}k^{ij}$. The operator
$\tr :T^{(m,n)}M\rightarrow T^{(m-1, n-1)}M$ is the trace with
respect to the metric, e.g. $\tr k =k^i_{\:i}$. For the sake of
simplicity in the second equation we also denote by $g$ and $k$ the
$(1,1)$-tensors with respect to the metric $g$ (i.e. $g=(g^i_{\:j})$,
$k=(k^i_{\:j})$ in the second equation). The linear differential operator
$\dv :C^\infty (T^{(m,n)}M)\rightarrow C^\infty
(T^{(m-1,n)}M)$ is the covariant divergence, defined by
\[\dv T:=\tr (\nabla T)\] 
where $T$ is a tensor field of $(m,n)$-type and $\nabla$ is the
Levi--Civit\'a covariant derivative of the metric $g$.
The smooth function $\rho : M\rightarrow \R$ is the energy-density, and
the smooth covector field $J\in C^\infty (T^*M)$ with $\vert
J\vert_g^2=\left< J, J\right> =J_iJ^i$ is interpreted as the
momentum-density of matter. 

Supposing the energy- and momentum-densities correspond to classical
non-dissipative matter sources or vacuum ($J=0$, $\rho=0$), the coupled
Einstein-equations can be used to evolve the initial data set $(M,g,k)$
into a (globally hyperbolic) smooth space-time $(N, h)$ where $N\cong
M\times\R$ and $M$ is a Cauchy surface in $N$; furthermore $h\vert_M=g$
and $k$ is the second fundamental form or extrinsic curvature of $M$ in
$(N, h)$ \cite{haw-ell} \cite{wal}. 

Also remember that the open oriented three-manifold $M$ has an {\it end}
$E\subset M$ if there is a compact set $C\subset M$ such that $M\setminus
C=E$ and $E\cong S_g\times(\R^+\setminus\{0\} )$ where $S_g$ is a compact,
oriented surface of genus $g$ and $\R^+=[0, \infty )$. An
initial value data set $(M,g,k)$ is called {\it asymptotically flat along
$E$} if $M$ has an end $E\cong S^2\times (\R^+\setminus\{ 0\} )$ and the
following asymptotical fall-off conditions hold for the complete metric
$g$ and the field $k$ ($r$ parameterizes $\R^+$ in $E$): 
\[\begin{array}{ll}
(g\vert_E)_{ij}=\delta_{ij}+O(1/r), 
& (k\vert_E)_{ij}=O(1/r^2),\\
& \\
\partial_l(g\vert_E)_{ij}=O(1/r^2), 
& \partial_l(k\vert_E)_{ij}=O(1/r^3),\\
& \\
\partial_l\partial_k(g\vert_E)_{ij}=O(1/r^3). &
\end{array}\]
Notice that the definition of a manifold with an end does not exclude the
possibility that $M$ still has a boundary, different from the end $E$
(strictly speaking, $E$ is not a boundary). But the boundary points are
added to $M$ because $C=M\setminus E$ is compact according
to our assumption. Consequently geodesic completeness of $g$ requires that
this extra boundary must be empty, in other words
$M$ is diffeomorphic to the punctured manifold $\widetilde{M}\setminus\{
y\}$ where $\widetilde{M}$ is a connected, compact, oriented 
three-manifold without boundary.

Finally, the {\it support} of a tensor field $T\in C^\infty (T^{(m,n)}W)$
is the closed set
\[\supp T:=\overline{\{ x\in W\:\vert\: T(x)\not=0\}}.\]
After this background material, we are in a position to prove our
theorem. The motivation is the following.
\section{The theorem}

It was proved by Witt \cite{wit} that every three-manifold with an end
admits a non-vacuum, asymptotically flat initial data set. For a typical
three-manifold, the resulting Cauchy developed space-time does not admit
maximal slices, however; i.e. there are no maximal space-like submanifolds
whose extrinsic curvature is identically zero.
One may raise the question: in what extent are these slices not maximal?
In other words what are the conditions on a Riemannian manifold $(M,g)$   
for its extrinsic curvature in the Cauchy development to be compactly
supported at least? We will answer this question in our theorem.

\vspace{0.1in}

{\bf Theorem} (Rigidity for non-vacuum initial data). {\it Let
$(M,g)$ be a connected, oriented, complete Riemannian three-manifold with
an end $E\cong S^2\times (\R^+\setminus\{ 0\} )$. Suppose the scalar
curvature $s_g$ of $g$ is non-positive everywhere and there is a
non-vacuum initial data set $(M,g,k)$ on it which is asymptotically flat
along the end $E$. Then $\supp k$ is non-compact.} 
\vspace{0.1in}

{\it Proof.} Since the scalar curvature is non-positive, the
set $\supp s_g$ consists of the closure of those points where $s_g$ is
negative. Then the first and third (in)equalities of (\ref{kenyszer}) show
that $\supp s_g\subseteq\supp k$ therefore if the scalar curvature is
negative everywhere the statement is trivially true, consequently we may
assume that $\supp s_g\subset M$. In the same fashion, since
$(M,g,k)$ is a non-vacuum data set, there is a point $x_0\in M$ such that
$\rho (x_0)\not= 0$. This yields $\supp\rho\not=\emptyset$. 
Being the scalar curvature non-positive, via the first and third
(in)equalities of (\ref{kenyszer}) again we have $k(x_0)\not=0$ i.e.,
$\supp\rho\subseteq\supp k$. Therefore if the energy density is supported
everywhere the theorem is again trivially valid consequently we may assume
$\supp\rho\subset M$. Consider
a subset $C\subset M$ such that $\supp s_g\subset C$ and $\supp\rho\subset
C$ and suppose the decomposition $M=C\cup E$ is valid where $E$
denotes the end of $M$. Consequently by the structure of $M$ we may assume
that $C$ is compact. This shows that there is a constant 
\[0<a:=\sup\limits_{x\in C}\left( -\vert k(x)\vert_g^2+\tr^2
k(x)\right) <\infty .\] 
Consider a triple $(M,g,\varphi g)$ where $\varphi : M\rightarrow\R$ is 
a $C^1$-function. This triple is a weak initial data set if it obeys the
constraint equations (here by ``weak'' we mean that the initial data set
in question is not smooth, only $C^k$ for some $k\in\N$):
\[\left\{ \begin{array}{ll}
 s_g + 6\varphi^2=16\pi\rho ,\\
\\
-2\:\dv (\varphi g)=-2\tr (\nabla (\varphi g))=-2\:\tr
(\dd\varphi\otimes g)=8\pi J,\\
\\
\rho\geq\vert J\vert_g\geq 0.
\end{array}\right. \]
In the second equation we have used the fact that $\nabla g=0$.
These (in)equalities can be combined into a first order partial
differential   inequality for the unknown function $\varphi$:
\begin{equation}
{1\over 4}(s_g + 6\varphi^2)\geq\vert\dd\varphi\vert_g  
\label{egyenlotlenseg}
\end{equation}
taking into account that $\vert\tr (\dd\varphi\otimes g
)\vert_g=\vert\dd\varphi\vert_g$. {\it Assume $\emptyset\not=\supp
k\subset M$ is compact i.e., the theorem is not true.} In this case we
construct a compactly supported function $\varphi$, out of the original
data $(M,g,k)$ such that $(M,g,\varphi g)$ is a weak initial data set. 
We achieve this in three steps. 

(i) {\it Construction of $\varphi$ in the compact interior of $M$}. 
Let us identify the end $E\subset M$ with $S^2\times
(\R^+\setminus\{ 0\} )$. By assumption $\supp k$ is compact in $M$
consequently there is an $R_1\in\R^+$ satisfying $S^2\times (R_1,\infty
)\not\subset\supp k$. Note that this is possible only if
$s_g\vert_{S^2\times (R_1,\infty )}=0$. We can take the choice
$C:=M\setminus (S^2\times (R_1, \infty ))$ for the compact set used in 
the definition of the constant $a$. We construct the function
$\varphi$ in $C$ as follows:
\[\varphi (x):=-\sqrt{a},\:\:\:\:\:x\in C.\]
In other words $\varphi$ is a constant negative function on $M$ except the
infinite tube $S^2\times (R_1, \infty )$. Note that with this function
(\ref{egyenlotlenseg}) is trivially satisfied in $C$ because $(M,g,k)$ is
an initial data set on $C$.
 
(ii) {\it Construction of $\varphi$ along an annulus in $E$.} 
Consider an inner point $x_0\in C\subset M$ where $\rho (x_0) >0$ and $k
(x_0)\not= 0$. There is an open (geodesic) ball $B_\varepsilon
(x_0)\subset M$ of radius $\varepsilon >0$ such that
$\rho\vert_{B_\varepsilon (x_0)}>0$ and
$k\vert_{B_\varepsilon (x_0)}\not= 0$. Consider the
annulus $U_\varepsilon :=\overline{B_\varepsilon
(x_0)\setminus B_{\varepsilon\over 2}(x_0)}\cong
S^2\times [{\varepsilon\over 2}, \varepsilon ]$. Take another
constant $R_1<R_2<\infty$ and the diffeomorphism
\[\beta : U_\varepsilon\longrightarrow
S^2\times [R_1,R_2],\:\:\:\:\:x_t=(p,t)\longmapsto 
\left( p\:,\: R_1+{2t-\varepsilon\over\varepsilon}
(R_2-R_1)\right) =(p,r)\]
where $p\in S^2$ and the point $x_t\in U_\varepsilon$ is identified
with $(p,t)\in S^2\times\left[{\varepsilon\over 2},
\varepsilon\right]$. Here $S^2\times [R_1, R_2]$ is also an annulus in
the tube $E$. By assumption $g$ is asymptotically flat i.e., the
function $\sqrt{g^{11}}\geq 0$ is bounded
consequently there is a constant
\[0<b:=\sup\limits_{x\in M}\sqrt{g^{11}(x)}<\infty\]
(here $x^1=r$). Choose a smooth function
$\psi :[R_1, R_2]\rightarrow\R^-$. Viewing it as a function on $S^2\times
[R_1, R_2]$ (i.e. a function depending only on $r$), one obtains the
estimate 
\begin{equation}
b\vert\psi '\vert\geq \vert
\sqrt{g^{11}}\psi '\vert =\vert\dd\psi\vert_g 
\label{becsles}
\end{equation}
where prime denotes differentiation with respect to $r$. Now we define
$\psi$ as follows:
\[\psi (\beta (x_t)):=\left\{\begin{array}{ll}
                   -\sqrt{a} & \mbox{if $t={\varepsilon\over 2}$,}\\
                    \mbox{arbitrary but the derivative of $\psi$
is small} & \mbox{if $t\in ({\varepsilon\over 2},\varepsilon )$},\\ 
                     0         & \mbox{if $t=\varepsilon$}.
                             \end{array}
                      \right.\]
In this definition the smallness of $\psi '$ means the following. Consider
a differentiable curve $\gamma : [{\varepsilon\over 2}, \varepsilon
]\rightarrow U_\varepsilon$ given by 
\[t\longmapsto x_t:=\left( \Theta_{\varepsilon\over 2}+A\sin
(R_2-R_1)t\:,\:\phi_{\varepsilon\over 2}+A\sin
(R_2-R_1)t\:,\: t\right) .\]
This is a high-speed curve because it oscillates rapidly inside
$B_\varepsilon (x_0)$. More precisely, for its speed 
$|\dot{\gamma}(t)|_g\sim R_2-R_1$
is valid (dot denotes differentiation with respect to $t$). We can take a
choice for the point $x_0$ and the amplitude $A$, the
initial phases $\Theta_{\varepsilon\over 2}$ and $\phi_{\varepsilon\over
2}$ of the curve $\gamma$ such that 
\[\dd\left(\sqrt{ -\vert k(x_t)\vert_g^2+\tr^2
k(x_t)}\right) (\dot{\gamma}(t))\sim -(R_2-R_1)<0\]
holds for each $t\in\left[ {\varepsilon\over 2} ,\varepsilon\right]$. Then
we suppose 
\begin{equation}
0\leq\psi '(\beta (x_t))\leq\min \left(-{\varepsilon\over
4(R_2-R_1)}\dd\left(\sqrt{ -\vert 
k(x_t)\vert_g^2+\tr^2 k(x_t)}\right) (\dot{\gamma}(t))\:,\:{16\pi\over
b}\rho (x_t)\right) .
\label{derivaltegyenlotlenseg}
\end{equation}
It is also clear that such a function exists if $R_2$ is suitable large: 
let $\psi$ be an arbitrary smooth, negative-valued function $\psi
:S^2\times [R_1, R_2]\rightarrow\R^-$ with initial value $\psi
(p,R_1)=\psi (\beta (x_{{\varepsilon\over 2}}))=-\sqrt{a}$. Suppose there is an
interval $[R, R+T]\subset [R_1, R_2]$ such that $\psi '$ obeys
(\ref{derivaltegyenlotlenseg}) but there is a constant $c>0$ with $\psi
'(p,r )\geq c$ if $r\in [R, R+T]$. This constant can be chosen to be
independent of $R_2-R_1$. In this case we can estimate for large $R_1$ and
$R_2$ as follows:
\[\psi (p, R_2)\geq-\sqrt{a}+{1\over 2}\int\limits_{R_1}^{R_2}\psi '(p, r
)\dd r\geq -\sqrt{a}+{1\over 2}\int\limits_R^{R+T}\psi '(p, r)\dd r \geq
-\sqrt{a} +{c\over 2}T.\]
In other words if $T$ that is, $R_2-R_1$ is sufficiently large we can
achieve that $\psi (p, R_2)=0$. We choose $\varphi$ on
$S^2\times [R_1,R_2]$ to be the $\psi$ just constructed.

It is not difficult to check that $\varphi$ obeys (\ref{egyenlotlenseg})
in $S^2\times [R_1, R_2]$. Indeed, by the definition of the constant $a$
we have 
\[\varphi(p, R_1 )=\varphi (\beta (x_{{\varepsilon\over
2}}))=-\sqrt{a}\leq -\sqrt{-\left\vert
k\left( x_{\varepsilon\over 2}\right)\right\vert_g^2+\tr^2
k\left( x_{\varepsilon\over 2}\right)}\:.\]
Taking suitable large $R_1$ and $R_2$, exploiting the decay of the
metric $g$ and using (\ref{derivaltegyenlotlenseg}) this implies that for
each $t\in\left[{\varepsilon\over 2}, \varepsilon\right]$ we have
\[\varphi (\beta (x_t))=\varphi 
(p,r)=-\sqrt{a}+\int\limits_{R_1}^r\dd\varphi 
(p, \varrho )(\beta '(x_\tau ))\dd\varrho =\]
\[-\sqrt{a}+\int\limits_{R_1}^r\varphi '(p,\varrho )\left( g_{11}(p,
\varrho )+A{\varepsilon\over
2}(g_{12}(p, \varrho )+g_{13}(p, \varrho ))\cos{\varepsilon\over
2}(\varrho +R_2)\right)\dd\varrho\leq\]
\[ -\sqrt{a}+2\int\limits_{R_1}^r\varphi '(p,\varrho )\dd\varrho 
=-\sqrt{a}+{4(R_2-R_1)\over\varepsilon}\int\limits_{{\varepsilon\over
2}}^t\varphi '(\beta (x_\tau ))\dd\tau\leq\]
\[ -\sqrt{-\left\vert
k\left( x_{{\varepsilon\over 2}}\right)\right\vert_g^2+\tr^2
k\left( x_{{\varepsilon\over 2}}\right)}
-\int\limits_{{\varepsilon\over 2}}^t\dd\left(
\sqrt{-\vert k(x_\tau )\vert_g^2+\tr^2
k(x_\tau )}\right) (\dot{\gamma}(\tau ))\dd\tau =-\sqrt{-\vert
k(x_t)\vert_g^2+\tr^2 k(x_t)}\:.\]
Consequently
\[\varphi^2 (\beta (x_t))\geq -\left\vert               
k(x_t)\right\vert_g^2+\tr^2k(x_t).\]
Therefore, since $s_g(\beta (x_t))=0$ and $0\geq s_g(x_t)$, we can
write
\[{1\over 4}\left( s_g(\beta (x_t))+6\varphi^2(\beta
(x_t))\right) ={3\over 2}\varphi^2 (\beta (x_t))\geq s_g(x_t)-\vert
k(x_t)\vert_g^2+\tr^2k(x_t) =16\pi\rho (x_t) .\]
Moreover, also by (\ref{derivaltegyenlotlenseg}), we have for the same
$x_t\in U_\varepsilon$ that $16\pi\rho (x_t)\geq b\varphi '(\beta (x_t))$.
This gives rise to our key inequality
\begin{equation}
{3\over 2}\varphi^2 (\beta (x_t))\geq b\varphi '(\beta (x_t))
\label{kulcsegyenlet}
\end{equation}
showing via (\ref{becsles}) that (\ref{egyenlotlenseg}) is again
satisfied in the annulus $S^2\times [R_1, R_2]$.

(iii) {\it Construction of $\varphi$ along the remaining part of
the infinitely long tube in $M$}. Finally, define 
\[\varphi (x):=0\:\:\:\:\:\mbox{if $x\in S^2\times [R_2, \infty )$.}\]
Again, (\ref{egyenlotlenseg}) is trivially valid. 

Consider the function $\varphi :M\rightarrow\R^-$ defined through
(i)-(iii). This is a continuous negative function on $M$ and is compactly
supported: it is equal to zero for all $r\geq R_2$ and equal to the
constant $-\sqrt{a}$ if $r\leq R_1$. Its derivative is also
compactly supported in $S^2\times [R_1, R_2]$ and is positive.
Moreover $\varphi$ can be adjusted to be $C^1$ on $M$ (note that 
$\varphi$ is smooth except the junction points): it is clearly $C^1$ at
$r=R_2$ by (\ref{kulcsegyenlet}). However, by exploiting the freedom 
in the construction of $\varphi$ in the inner points of the annulus, 
we can deform it to be $C^1$ at $r=R_1$ as well
(i.e., we may assume that $\varphi '(p, r)\rightarrow 0$ as $r\rightarrow
R_1$). In this way we have constructed a weak $C^1$
initial data set $(M,g,\varphi g)$ (with a little effort we could smooth
this data but we do not need this).

The compactly supported $\varphi$ depends nontrivially only on $r$ with
$(p,r)=\beta (x_t)\in S^2\times [R_1, R_2]$ and satisfies the ordinary
differential inequality (\ref{kulcsegyenlet}). Now we demonstrate that it
is impossible. Dividing by $\varphi '^2$ and taking reciprocies in
(\ref{kulcsegyenlet}) we get
\[\left({\varphi '\over\varphi}\right)^2\leq {3\varphi '\over 2b},\]
which is nothing but
\[-\sqrt{{3\varphi '\over 2b}}\leq {\varphi 
'\over\varphi}\leq\sqrt{{3\varphi '\over 2b}}.\]
By integrating the left inequality from $R_1$ to $r<R_2$ we arrive at the
following estimate:
\[\log\sqrt{a}-\sqrt{3\over 2b}\int\limits_{R_1}^{R_2}\sqrt{\varphi '(p,
\varrho )}\:\dd\varrho\leq\log (-\varphi (p, r)).\]
At this point we have used the inequality
\[0<\int\limits_{R_1}^r\sqrt{\varphi '(p, \varrho )}\:\dd\varrho\leq
\int\limits_{R_1}^{R_2}\sqrt{\varphi '(p, \varrho )}\:\dd\varrho <\infty
\]
for the non-negaitve function $\varphi '$. This shows that the logarithm
of $\varphi$ is bounded from below. However being $\varphi$ compactly
supported, $\log (-\varphi (p, r))$ is unbounded, as $r$
approaches $R_2$. Consequently the last but one inequality 
shows a contradiction yielding our original assumption, that $\supp k$ is
compact, was wrong. We finished the proof. $\Diamond$ 
\vspace{0.1in}

\noindent {\it Remarks.} 1.We would like to summarize here how the
original initial data $(M,g,k)$ was used in the construction because
apparently its behaviour has been taken into account only in a particular
small ball $B_\varepsilon (x_0)$. But in fact the construction
is sensitive for the global characteristics of the original initial
data. In step (i) we considered $(M,g,k)$ in the whole interior $C$ by
exploiting the existence of the constant $a$ which is in some sense the
maximum of $k$ in the whole compact $C$. This enabled us to ``pump up''
the original initial data in $C$ into a standard one which
corresponds to the extremal point(s) of the original extrinsic
curvature in some sense. Concerning part (iii), we have seen in
the beginning of the proof that the only interesting possibility for our
would-be initial data with compactly supported extrinsic curvature was the
case where both the scalar curvature and energy-density were compactly
supported. Consequently all fields in the initial data vanish along the
tube for very large $r$ yielding the hypothetical initial data did
not carry ``information'' along an infinitely long part of the end
$E$. This is in accordance with the fact that our adjusted universal
initial data $(M,g,\varphi g)$ was also trivial on this portion.   
Finally, part (ii) which is the descending regime, is nothing but a
magnification of the behaviour of $(M,g,k)$ in a small ball where matter
is present via the diffeomorphism $\beta$. Indeed this small ball is
responsible for the details of the fall-off of $\varphi$ (we could have 
used equally well any other ball) however the fact that this
function can vanish within a finite distance, is again guaranteed by the
global properties of the original would-be initial data set: namely the
only interesting case was when all fields were compactly supported.

2. Note that even if $\supp k$ is non-compact the
non-vacuum data $(M,g,k)$ may be asymptotically flat, as it is shown by
Witt \cite{wit} who constructs non-vacuum, asymptotically flat initial
data for every three-manifold with an end. But the above theorem is sharp
in the following sense. If we allow for a Riemannian manifold $(M,g)$ to
have positive scalar curvature in a suitable region in $M$, it is 
possible to construct non-vacuum asymptotically flat initial data with
compactly supported second fundamental form. An example is the
Tolman--Bondi solution. This is because in this case the key inequality
(\ref{kulcsegyenlet}) can be written in the form
\[{1\over 4}( s_g + 6\varphi^2)\geq b\varphi '\]
with $s_g>0$ in the positive scalar curvature regime and it may have
compactly supported solutions. But if $s_g$ is still negative somewhere,
then $k$ is non-zero in that point, consequently the initial surface is not a
maximal slice in this case. 

3. Notice that the above considerations do not remain valid for {\it
vacuum initial data}. For example, the Schwarzschild space-time has
initial data with non-positive scalar curvature (namely it is identically
zero) but the extrinsic curvature of the initial surface is compactly
supported (namely identically zero i.e., the initial surface is a
maximal slice). We conjecture that the analogue of the above theorem for
vacuum initial data is the following: if $(M,g)$ is an asymptotically flat
three-manifold with {\it some}where negative scalar curvature, then any
vacuum initial data set $(M,g,k)$ cannot be asymptotically flat (i.e., the
extrinsic curvature cannot decay at the required rate). But in this
moment we are unable to prove this.

\section{Concluding Remarks}
In the previous section we have studied gravitational initial data   
from a general point of view. We have found that in the case of
non-positive scalar curvature, the behaviour of the extrinsic curvature
becomes very ``rigid'': for open manifolds, the fall-off of the extrinsic
curvature cannot be arbitrary. 

The negativity of the scalar curvature becomes important by an early
general result of Kazdan and Warner\cite{kaz-war}: 
\vspace{0.1in}

{\bf Theorem} (Kazdan--Warner). {\it Let $W$ be a compact manifold with
$\dim W\geq 3$, and $f: W\rightarrow\R$ be a smooth function on it such
that there is a point $x\in W$ obeying $f(x)<0$. Then there is a smooth
Riemannian metric $h$ on $W$ such that $s_h=f$ i.e. whose scalar
curvature is the prescribed function $f$.} $\Diamond$
\vspace{0.1in}

\noindent The theorem demonstrates that a compact manifold of
sufficiently large dimension always can be endowed with a metric with
somewhere negative scalar curvature. This shows, taking into account the
constraint equations, that it is relatively easy to construct initial data
with somewhere non-zero extrinsic curvature.

The classical results of Gromow--Lawson \cite{gro-law} and Schoen-Yau
\cite{sch}, however, show that closed three-manifolds whose prime
decomposition contains a $K(\pi ,1)$ factor (this implies such manifolds 
have infinite fundamental groups) do not carry any metric with positive
scalar curvature. Consequently initial data with positive scalar
curvature must be rare at least in the compact case.

If the compact $\widetilde{M}$ does not have positive scalar 
curvature, the punctured, open manifold $M=\widetilde{M}\setminus\{ y\}$ 
of this type, which is nothing but a manifold with an end, does not have a
metric with non-negative scalar curvature, too. Consequently these
punctured manifolds do not admit non-vacuum, asymptotically flat initial
data with identically zero extrinsic curvature. Furthermore if the scalar
curvature is everywhere non-positive then this extrinsic curvature has
non-compact support, as we have seen.

These results are quite surprising because all the fields in question are
defined in the class of smooth functions so one would expect
that initial data can be altered {\it locally} in a non-trivial way. In
other words we have reduced the local degrees of freedom of
the gravitational field in some sense.


\begin{thebibliography}{99}

\bibitem{bes} Besse, A.: Einstein Manifolds, Springer-Verlag,
Berlin (1987);

\bibitem{bru} Bruhat, Y.: {\it The Cauchy Problem},
in: Gravitation (ed.: L. Witten), 130-168, John Wiley \& Sons, Inc., New
York, London (1962);

\bibitem{cho-ger} Choquet-Bruhat, Y., Geroch,
R.P.: {\it Global Aspects of the Cauchy Problem in General Relativity},
Commun. Math. Phys. {\bf 14}, 329-335 (1969);

\bibitem{cho-ise-mon}
Choquet-Bruhat, Y., Isenberg, J., Moncrief, V.: {\it Solution of
constraints for Einstein equations}, C.R. Acad. Sci. Paris,
S\'er. I {\bf 315}, 349-355 (1992);

\bibitem{cho-ise-yor} Choquet-Bruhat,
Y., Isenberg, J., York Jr., J.W.: {\it Einstein constraints on
asymptotically Euclidean manifolds}, Phys. Rev. {\bf D61},
084034-1-084034-20 (2000);
 
\bibitem{cho-yor} Choquet-Bruhat, Y., York Jr.,
J. W.: {\it The Cauchy Problem}, in: General Relativity and Gravitation
(ed.: A. Held), Vol. {\bf 1}, 99-172, Plenum Press, New York (1980);

\bibitem{chr} Christodoulou, D.,
Klainerman, S.: The Global Nonlinear Stability of the Minkowski
Space-Time, Princeton Univ. Press, Princeton, New Jersey (1993);

\bibitem{mar-fis} Fisher, A.E., Marsden, J.E.: {\it
Initial Value Problem and the Dynamical Formulation of General
Relativity}, in: General Relativity (ed.: S.W. Hawking and W. Israel),
138-211, Cambridge University Press, Cambridge (1979);

\bibitem{gro-law} Gromow, M., Lawson, H.B.: {\it
Positive Scalar Curvature and the Dirac Operator on Complete Riemannian
Manifolds}, Publ. Math. IHES {\bf 58}, 83-196 (1983);

\bibitem{haw-ell} Hawking, S.W., Ellis, G.F.R.: The
Large Scale Structure of Space-Time, Cambridge University Press, Cambridge
(1973);

\bibitem{hit} Hitchin, N.: {\it Harmonic Spinors},
Adv. Math. {\bf 14}, 1-55 (1974);

\bibitem{ise} Isenberg, J.: {\it Constant mean curvature
solutions of the Einstein constraints on closed manifolds},
Class. Quant. Grav. {\bf 12}, 2249-2274 (1995);

\bibitem{ise-mon} Isenberg, J., Moncrief,
V.: {\it A set of constant mean curvature solutions of the Einstein
constraint equations on closed manifolds}, Class. Quant. Grav. {\bf 13},
1819-1842 (1996);

\bibitem{kaz-war} Kazdan, J.L., Warner, F.W.: {\it
Scalar Curvature and Conformal Deformation of Riemannian Structure},
Journ. Diff. Geom. {\bf 10}, 113-134 (1975);

\bibitem{lic1} Lichnerowicz, A.: {\it L'integration des
Equations de la Gravitation Relativiste et le Probleme des $n$ Corps},
J. Math. Pures Appl. {\bf 23} 37-63 (1944);
 
\bibitem{lic2} Lichnerowicz, A.: {\it Spineurs
Harmoniques} C.R. Acad. Sci. Paris {\bf 257}, 7-9 (1963);

\bibitem{sch} Schoen, R.: {\it Minimal Surfaces and Positive
Scalar Curvature}, in: Proceedings of the International Congress of
Mathematics (Warszawa, 1983), pp. 575-578, PWN, Warszawa (1984);

\bibitem{wal} Wald, R.M.: General Relativity, University of
Chicago Press, Chicago (1984);

\bibitem{wit} Witt, D.M.: {\it Vacuum Spacetimes that Admit no
Maximal Slice}, Phys. Rev. Lett. {\bf 57}, 1386-1389 (1986).

\end{thebibliography}
\end{document}